%% file: main.tex
\newif\ifdraft
\draftfalse

\newif\ifcameraready
\camerareadytrue

\documentclass[camera,letterpaper,nomarginnotes,nonarrowgutter]{jpaper}
\newcounter{version}
\input{packages}

\input{glossary}
\input{macros}
\usepackage{lscape}

\newcites{Mod}{Module References}

\def\BibTeX{{\rm B\kern-.05em{\sc i\kern-.025em b}\kern-.08em
    T\kern-.1667em\lower.7ex\hbox{E}\kern-.125emX}}
    
\newcommand{\versionnum}[0]{1.06}

\ifcameraready
    \fancypagestyle{plain}{
    \fancyhead{}
    }
\else
    \fancypagestyle{firstpage}
    {
    }
\fi

\makeatletter
\def\bstctlcite{\@ifnextchar[{\@bstctlcite}{\@bstctlcite[@auxout]}}
\def\@bstctlcite[#1]#2{\@bsphack
  \@for\@citeb:=#2\do{%
    \edef\@citeb{\expandafter\@firstofone\@citeb}%
    \if@filesw\immediate\write\csname #1\endcsname{\string\citation{\@citeb}}\fi}%
  \@esphack}
\makeatother 

\begin{document}
\title{In-DRAM Signature Generation Using Simultaneous Multiple-Row Activation: An Experimental Study of Off-The-Shelf DRAM Chips\vspace{-2mm}}

\newcommand{\affilETH}[0]{\textsuperscript{\S}}
\newcommand{\affilTOBB}[0]{\textsuperscript{$\dagger$}}
\newcommand{\affilSharjah}[0]{\textsuperscript{$\ddagger$}}
\author{
{Umut~Ba\c{s}er\affilETH\affilTOBB}~~~~
{{\.I}smail~Emir~Y{\"u}ksel\affilETH}~~~~%
{F.~Nisa~Bostanc{\i}\affilETH}~~~~%
{Konstantinos~Sgouras\affilETH}~~~~%
{Ataberk~Olgun\affilETH}~~~~\\%
{Emre~Hakan~Demirli\affilETH\affilTOBB}~~~~%
{Zhiheng~Yue\affilETH}~~~~%
{Harsh~Songara\affilETH}~~~~%
{O\u{g}uz~Ergin\affilSharjah\affilTOBB}~~~~%
{Onur~Mutlu\affilETH}
\vspace{-3mm}\\\\
{\affilETH ETH Z{\"u}rich\qquad{} \affilTOBB TOBB ETÜ\qquad{} \affilSharjah University of Sharjah}
\vspace{-3.5mm}
}

\maketitle
\ifcameraready
    \thispagestyle{plain}
\else
    \thispagestyle{firstpage}
\fi

\pagestyle{plain}
\setcounter{version}{3}

\input{sections/00_abstract}
\glsresetall
\glsresetall
\input{sections/01_introduction}
\glsresetall
\input{sections/02_background}
\input{sections/03_methodology}
\input{sections/04_simra-puf}
\input{sections/05_related-work}
\input{sections/06_discussion}

\input{sections/ack.tex}

\balance
\bibliographystyle{unsrt}
\bibliography{refs}
\vfill

\makeatletter
\c@enumiv=0\relax
\makeatother

\input{sections/07_appendix}
\clearpage
\emergencystretch=3em
\bibliographystyleMod{unsrt}
\bibliographyMod{module_refs}
\vfill

\end{document}

%% file: packages.tex
\usepackage[bookmarks=true,breaklinks=true,colorlinks,linkcolor=black,citecolor=NavyBlue,urlcolor=blue]{hyperref}
\usepackage{xurl}
\usepackage{algorithmic}
\usepackage{graphicx}
\usepackage{textcomp}
\usepackage{xcolor}
\usepackage[flushleft]{threeparttable}

\usepackage{titlecaps}

\usepackage{balance}
\usepackage{textcomp}
\usepackage{pifont}
\usepackage{booktabs}
\usepackage{glossaries}
\setkeys{glslink}{hyper=false}
\usepackage{setspace}
\usepackage{listings}
\usepackage[linesnumbered]{algorithm2e}
\usepackage{siunitx}
\usepackage{subcaption}
\usepackage{tikz}
\usepackage[dvipsnames]{xcolor}
\usepackage{multirow}
\usepackage{enumitem}
\usepackage[utf8]{inputenc}

\usepackage{titlesec}

\usepackage{interval} %sec5, interval
\usepackage{duckuments} %duck figures
\usepackage{algorithmic} %algoritmic

\usepackage[us,12hr]{datetime}
\usepackage[en-GB, useregional=numeric]{datetime2}
\usepackage{fourier-orns}
\usepackage{fancyhdr}
\usepackage{pgfornament}
\usetikzlibrary{calc}
\usepackage{url}
\usepackage{yfonts}
\usepackage[sort,compress]{cite}

\usepackage{mathptmx} % This is Times font

\usepackage[resetlabels]{multibib}
\usepackage{pdflscape}

%% file: glossary.tex
\newcommand{\tras}[0]{t_{RAS}}
\newcommand{\trp}[0]{t_{RP}}
\newcommand{\trc}[0]{t_{RC}}
\newcommand{\trefi}[0]{t_{REFI}}
\newcommand{\trefw}[0]{t_{REFW}}
\newcommand{\trfc}[0]{t_{RFC}}
\newcommand{\trrd}[0]{t_{RRD}}

\newcommand{\nrg}[0]{$N_{RG}$}

\newcommand{\act}[0]{\texttt{ACT}}
\newcommand{\pre}[0]{\texttt{PRE}}
\newcommand{\refresh}[0]{REF}
\newcommand{\wri}[0]{\texttt{WR}}
\newcommand{\rd}[0]{\texttt{RD}}

\newacronym{acmin}{$AC_{min}$}{the minimum number of total aggressor row activations to cause at least one bitflip}

\newcommand{\apa}[0]{\texttt{APA}}

\newcommand{\vddh}[0]{\texttt{$V_{DD}/2$}}

\newcommand{\pum}[0]{PUM}
\newcommand{\pim}[0]{PIM}
\newcommand{\pnm}[0]{PNM}

\newcommand{\pud}[0]{{PUD}}

\usepackage[shortcuts]{extdash}

\newacronym{iqr}{$IQR$}{inter-quartile range}
\newacronym{act}{\act{}}{activate}
\newacronym{pre}{\pre{}}{precharge}
\newacronym{ref}{\refresh{}}{refresh}
\newacronym{wr}{\wri{}}{write}
\newacronym{rd}{\rd{}}{read}
\newacronym{pim}{\pim{}}{Processing-In-Memory}
\newacronym{pnm}{\pnm{}}{Processing-Near-Memory}

\newacronym{pum}{\pum{}}{Processing-Using-Memory}
\newacronym{pud}{\pud{}}{Processing-Using-DRAM}
\newacronym{apa}{\apa{}}{\act{} $\rightarrow$ \pre{} $\rightarrow$ \act{}}
\newacronym{jedec}{JEDEC}{Joint Electron Device Engineering Council}
\newacronym{nrg}{\nrg{}}{a group of N rows activated simultaneously with \apa{} sequence as a row group}

\newacronym{trefw}{$\trefw$}{refresh window}
\newacronym{tras}{$\tras$}{the latency of sensing the row's data and fully restoring a DRAM cell's charge}
\newacronym{trp}{$\trp$}{the latency of de-asserting a wordline and precharging the bitlines to \vddh{}}
\newacronym{trc}{$\trc$}{the minimum time needed between two consecutive row activations targeting the same bank}
\newacronym{trefi}{$\trefi$}{refresh interval}
\newacronym{trfc}{$\trfc$}{refresh latency}
\newacronym{trrd}{$\trrd$}{the minimum time needed between two consecutive row activations targeting the same rank}
\newacronym{puf}{PUF}{physical unclonable function}
\newacronym{trn}{TRN}{true random number}

%% file: macros.tex
\definecolor{ksgo}{rgb}{0.580, 0.050, 0.211}
\definecolor{ao}{rgb}{0.007, 0.520, 0.867}
\definecolor{moegi}{rgb}{0.357, 0.537, 0.188}
\definecolor{jl}{rgb}{1.0, 0.2, 0.8}
\definecolor{brown(web)}{rgb}{0.65, 0.16, 0.16}
\definecolor{bisque}{rgb}{1.0, 0.89, 0.77}
\definecolor{nbs}{rgb}{0.88, 0.07, 0.37}
\definecolor{yt}{rgb}{0.58, 0.44, 0.86}
\definecolor{iy}{rgb}{0.0, 0.36, 0.05}
\definecolor{ouscolor}{rgb}{0.0, 0.2, 0.4}
\newcommand{\xxx}[1]{\param{XXX}} % to highlight hardcoded numbers

\newcommand{\ignore}[1]{}

\ifdraft
    \usepackage[colorinlistoftodos,prependcaption,textsize=tiny]{todonotes}
    \usepackage{color,soul}
    \paperwidth=\dimexpr \paperwidth + 4cm\relax
    \oddsidemargin=\dimexpr\oddsidemargin + 2cm\relax
    \evensidemargin=\dimexpr\evensidemargin + 2cm\relax
    \marginparwidth=\dimexpr \marginparwidth + 2cm\relax
    \newcommand{\param}[1]{\textcolor{red}{#1}} % to highlight hardcoded numbers
    
    \newcommand{\mscomment}[1]{\todo[size=\scriptsize, linecolor=orange, bordercolor=orange, backgroundcolor=white]{\textcolor{red}{\textbf{@MS:} #1}}}

    \newcommand{\atbcomment}[1]{\todo[size=\scriptsize, linecolor=orange, bordercolor=orange, backgroundcolor=white]{\textcolor{ao}{\textbf{@atb:} #1}}}

    \newcommand{\ouscomment}[1]{\todo[size=\scriptsize, linecolor=orange, bordercolor=orange, backgroundcolor=white]{\textcolor{ouscolor}{\textbf{@ous:} #1}}}

    \newcommand{\yctcomment}[1]{\todo[size=\scriptsize, linecolor=orange, bordercolor=orange, backgroundcolor=white]{\textcolor{yt}{\textbf{@yct:} #1}}}

    \newcommand{\gfcomment}[1]{\todo[size=\scriptsize, linecolor=orange, bordercolor=orange, backgroundcolor=white]{\textcolor{blue}{\textbf{@gf:} #1}}}

    \newcommand{\nbcomment}[1]{\todo[size=\scriptsize, linecolor=orange, bordercolor=orange, backgroundcolor=white]{\textcolor{nbs}{\textbf{@nb:} #1}}}

    \newcommand{\hluocomment}[1]{\todo[size=\scriptsize, linecolor=orange, bordercolor=orange, backgroundcolor=white]{\textcolor{moegi}{\textbf{@hluo:} #1}}}

    \newcommand{\ieyinline}[1]{{\color{iy}{\textbf{\hl{[@iey:}} \textit{\hl{#1}}\textbf{]}}}}
    \newcommand{\ieycomment}[1]{\todo[size=\scriptsize, linecolor=orange, bordercolor=orange, backgroundcolor=white]{\textcolor{iy}{\textbf{@iey:} #1}}}

    \newcommand{\omcomment}[1]{\todo[size=\scriptsize, linecolor=orange, bordercolor=orange, backgroundcolor=white]{\textcolor{teal}{\textbf{@ste:} #1}}}
    \newcommand{\ominline}[1]{{\textcolor{teal}{\textbf{[@ste:} #1\textbf{]}}}}

\else
    
    \newcommand{\param}[1]{\textcolor{black}{#1}} % to highlight hardcoded numbers
    \newcommand{\ksgcomment}[1]{}
    \newcommand{\agyinline}[1]{}

    \newcommand{\mscomment}[1]{}

    \newcommand{\atbcomment}[1]{}

    \newcommand{\yctcomment}[1]{}

    \newcommand{\gfcomment}[1]{}

    \newcommand{\nbcomment}[1]{}

    \newcommand{\ouscomment}[1]{}

    \newcommand{\hluocomment}[1]{}

    \newcommand{\ieycomment}[1]{}
    \newcommand{\ieyinline}[1]{}

    \newcommand{\omcomment}[1]{}
    \newcommand{\ominline}[1]{}

\fi

\newcommand*\obsv[1]{%
\refstepcounter{obs}
\noindent\textit{\underline{Observation~\theobs.}} \emph{#1}}

\newcounter{obs}
\newcounter{tkw}
\newcommand{\changedText}[1]{%
  \ifcameraready
    #1%
  \else
    \textcolor{blue}{#1}%
  \fi
}

\newcommand{\actTwoInterJac}{3.98\%}
\newcommand{\actFourInterJac}{2.37\%}
\newcommand{\actEightInterJac}{3.44\%}
\newcommand{\actSixteenInterJac}{2.92\%}
\newcommand{\actThirtyTwoInterJac}{3.24\%}

\newcommand{\actTwoIntraJac}{89.02\%}
\newcommand{\actFourIntraJac}{89.81\%}
\newcommand{\actEightIntraJac}{93.03\%}
\newcommand{\actSixteenIntraJac}{94.06\%}
\newcommand{\actThirtyTwoIntraJac}{94.86\%}

\newcommand{\fracIntraJac}{95.62\%} %95.44 previous 9 modules

\newcommand{\fracGapTwoRow}{6.60\%}      
\newcommand{\fracGapThirtyTwoRow}{0.76\%}

\newcommand{\chipCount}{112}
\newcommand{\keyObservationCount}{8}
\newcommand{\moduleCount}{10}
\newcommand{\fracLatencyDiff}{5.75\%}
\newcommand{\fracLatencyDiffThirtyTwoRow}{3.84$\times$}

\newcommand{\dramPUFCiteBlock}{\cite{
kim2018dram,
najafi2021,
talukder2019prelatpuf,
najafi2025epuf,
miskelly2020fastdrampuf,
Hashemian:2015,
Sutar2016,
Xiong2016,
schaller2019,
Keller2014,
zheng2021implementation,
liu2014,
kumari2018rapid,
achievingerrorfree,
9923819,
orosa2021codic,
schaller:2017,
FISCHER2025,
anagnostopoulos2018intrinsic,
li2023fphammer,
Tehranipoor:2015,
tehranipoor2016dram}}

\newcommand{\modifiedPumCiteBlock}{\cite{\modifiedPudCiteBlock}}
\newcommand{\modifiedPudCiteBlock}
{seshadri2017ambit, seshadri2019dram, drisa, seshadri2013rowclone, seshadri2015fast, elp2im, xinroc, seshadri2016buddy, seshadri2017simple, scope, pluto, manning2018apparatuses, zawodny2018apparatuses, dracc, flexidram, deng2019lacc, hajinazar2021simdram, oliveira2022methodologies,oliveira2022accelerating, deoliveira2024mimdram, geraldo2025proteus, shin2025prada, seshadri2015gather, seshadri2016processingusingmemoryparadigm,wang2020figaro, graphide, sisa, sutradhar2020, sutradhar2022, dramcam, tokuda2026clutch}

\newcommand{\sramPumCiteBlock}{\cite{kang.icassp14, aga.hpca17, eckert2018neural, fujiki2019duality, agrawal2018x, al2020towards, jeloka201628, jiang2020c3sram, kang2015energy, kim2021colonnade, nag2019gencache, si2019dual, simon2020blade, wang2019bit, wang2023infinity, Ramanathan2020-lut-sram, sharma2025flexdcim}}
\newcommand{\flashPumCiteBlock}{\cite{flashcosmos,gao2021parabit,
ciphermatch, conduit, aif,kang2021s, chen2024ares,crossbit, wong2024tcam, anvil,chun-hotstorage-2022,choi2020flash,wang2018three,shim-ieeejetcas-2022,lee20223d}}
\newcommand{\emergingPumCiteBlock}{\cite{li2016pinatubo, 
shafiee2016isaac,
kvatinsky.tcasii14, kvatinsky.iccd11, kvatinsky.tvlsi14, gaillardon2016plim, %bhattacharjee2017revamp, 
hamdioui2015memristor, xie2015fast, hamdioui2017myth, 
truong2021racer, 
ma20232, slesazeck20192tnc, 
song2017pipelayer, rakin2018-tgan, 
zha2020-hyperap-rram, song2018graphr}}

\newcommand{\jaccardDRAMCiteBlock}{\cite{Xiong2016, kim2018dram, orosa2021codic, najafi2021, schaller:2017,FISCHER2025, schaller2019}}

\newcommand{\apaReferences}{
gao2019computedram, olgun2021quac, 9923819,  olgun2022pidram,olgun2022drambender,yuksel2025in-dram, yuksel2024functionally, yuksel2024simultaneous,yuksel2024pulsarsimultaneousmanyrowactivation, yuksel2025pudhammer,yaglikci2022hira
}

\newcommand{\figref}[1]{Fig.~\ref{#1}}

\newcommand{\secref}[1]{§\ref{#1}}

\ifcameraready

    \newcommand{\nbcr}[2]{\ifnum#1>-1\textcolor{black}{#2}\else{#2}\fi}
    \newcommand{\omcr}[2]{\ifnum#1>-1\textcolor{black}{#2}\else{#2}\fi}
    \newcommand{\omcrcomment}[1]{}
    \newcommand{\crdiscussion}[2]{}{}

    \newcommand{\ieycrcomment}[1]{}
    \newcommand{\atbcrcomment}[1]{}
    \newcommand{\agycrcomment}[1]{}
\else
\usepackage[colorinlistoftodos,prependcaption,textsize=scriptsize]{todonotes} % For margin notes
    \paperwidth=\dimexpr \paperwidth + 4cm\relax
    \oddsidemargin=\dimexpr\oddsidemargin + 2cm\relax
    \evensidemargin=\dimexpr\evensidemargin + 2cm\relax
    \marginparwidth=\dimexpr \marginparwidth + 2cm\relax

    \newcommand{\nbcr}[2]{\ifnum#1=\value{version}\textcolor{nbs}{#2}\else{#2}\fi}

    \newcommand{\ieycrcomment}[1]{\todo[linecolor=orange, bordercolor=orange, backgroundcolor=white]{\textcolor{iy}{\textbf{@Ismail:} #1}}}

    \newcommand{\atbcrcomment}[1]{\todo[linecolor=brown, bordercolor=brown, backgroundcolor=white]{\textcolor{ao}{\textbf{@Atb:} #1}}}

    \newcommand{\agycrcomment}[1]{\todo[size=\scriptsize, linecolor=orange, bordercolor=orange, backgroundcolor=white]{\textcolor{gfored}{\textbf{@gy:} #1}}}
    
    \newcommand{\crdiscussion}[2]{\omcrcomment{#1\\\textcolor{blue}{\textbf{@Ismail:}#2}}}
    \newcommand{\omcr}[2]{\ifnum#1=\value{version}\textcolor{red}{#2}\else{#2}\fi}

    \newcommand{\omcrcomment}[1]{\todo[linecolor=orange, bordercolor=orange, backgroundcolor=white]{\textcolor{red}{\textbf{@Onur:} #1}}}

\fi

\makeatletter
\g@addto@macro{\normalsize}{%
  \setlength{\abovedisplayskip}{1pt plus 0.5pt minus 1pt}
  \setlength{\belowdisplayskip}{1pt plus 0.5pt minus 1pt}
  \setlength{\abovedisplayshortskip}{0pt}
  \setlength{\belowdisplayshortskip}{0pt}
  \setlength{\intextsep}{0pt plus 1pt minus 1pt}
  \setlength{\textfloatsep}{3pt plus 1pt minus 1pt}
  \setlength{\skip\footins}{3pt plus 1pt minus 1pt}
  \setlength{\abovecaptionskip}{1pt plus 0pt minus 0pt}}
\makeatother

\usepackage{titlesec}
\titlespacing\section{0pt}{2pt plus 1pt minus 1pt}{1pt plus 1pt minus 1pt}
\titlespacing\subsection{0pt}{2pt plus 1pt minus 1pt}{1pt plus 1pt minus 1pt}
\titlespacing\subsubsection{0pt}{2pt plus 1pt minus 1pt}{1pt plus 1pt minus 1pt}

%% file: sections/00_abstract.tex
\begin{abstract}
We experimentally demonstrate that it is possible to generate unique, repeatable, and device-specific signatures suitable for use as Physical Unclonable Function (PUF) responses in commercial off-the-shelf (COTS) DRAM chips by leveraging simultaneous multiple-row activation (SiMRA). Based on a rigorous experimental characterization of \chipCount{} modern DDR4 DRAM chips (from \moduleCount{} modules), we introduce SiMRA-PUF, the first DRAM-based PUF that uses SiMRA-generated signatures as PUF responses. We analyze SiMRA-PUF in terms of reliability, uniqueness, and evaluation latency for varying numbers of simultaneously activated DRAM rows (i.e., 2, 4, 8, 16, and 32), DRAM chip density \& die revision, and evaluate how temperature affects the similarity of SiMRA-generated responses. Among our \keyObservationCount{} key experimental observations, we highlight two major results. First, SiMRA-PUF provides average intra-Jaccard indices of \actTwoIntraJac, \actFourIntraJac, \actEightIntraJac, \actSixteenIntraJac, and \actThirtyTwoIntraJac, and average inter-Jaccard indices of \actTwoInterJac, \actFourInterJac, \actEightInterJac, \actSixteenInterJac, and \actThirtyTwoInterJac{} for 2-, 4-, 8-, 16-, and 32-row activations, respectively, showing that SiMRA-generated signatures are both repeatable within a device and unique across devices. Second, 2-row activation-based SiMRA-PUF provides \fracLatencyDiff{} lower evaluation latency than the state-of-the-art DRAM-based PUF. We open-source our infrastructure and datasets at \url{https://github.com/CMU-SAFARI/SiMRA-PUF}.
\end{abstract}

%% file: sections/01_introduction.tex
\section{Introduction}
\label{sec:introduction}
Physically Unclonable Functions (PUFs) map a unique input (i.e., challenge) to a unique, repeatable, and device-specific signature (i.e., response). By providing repeatable and device-specific responses, PUFs are suitable for low-cost authentication protocols~\cite{Che2015,Hammouri2008,Majzoobi2012,Rostami2014, devadas2008, chatterjee2019}, key generation applications~\cite{Roel2012,Paral2011,Yu2012, Delvaux2015, rahman2016, korenda2019}, intellectual property protection~\cite{guajardo2007fpga,guajardo2008brand,zheng2014digital, kumar2008butterfly,zhang2015, zhang2013, guo2018, bordel2019digital, xue2025}, hardware obfuscation~\cite{khaleghi2018hardware,wendt2014hardware}, and the prevention of hardware trojan embedding~\cite{mobaraki2018novel}.

DRAM-based PUFs~\dramPUFCiteBlock{} leverage the characteristics of DRAM circuitry to generate unique signatures. DRAM-based PUFs are promising PUF designs for two key reasons: 1) DRAM is the dominant memory technology used as main memory in almost all computing systems~\cite{mutlu2013memory, mutlu2025memorycentric, mutlu2020modern}, ranging from edge devices to supercomputers, and 2) DRAM's high density provides a large challenge-response space for PUFs.

Recent works~\cite{9923819, gao2019computedram, yuksel2024pulsarsimultaneousmanyrowactivation, yuksel2024simultaneous, yuksel2024functionally, olgun2021quac, yuksel2025in-dram, mutlu2024memory} experimentally demonstrate a new phenomenon in DRAM: simultaneous multiple-row activation (SiMRA). By carefully engineering a sequence of DRAM commands,
SiMRA enables the simultaneous activation of up to 32 rows within a single subarray of real DRAM chips. Prior works~\cite{yuksel2024simultaneous, yuksel2024functionally, yuksel2025in-dram,yuksel2024pulsarsimultaneousmanyrowactivation, olgun2021quac, gao2019computedram, 9923819, mutlu2024memory} show that SiMRA can be leveraged to perform many in-DRAM operations (e.g., bulk bitwise operations).

We hypothesize that SiMRA could be used to generate PUF responses because manufacturing process variations could introduce device-specific outcomes due to the charge-sharing process when multiple DRAM rows are simultaneously activated. Therefore, SiMRA could potentially be leveraged as a fast and low-overhead PUF substrate.

Our \emph{goal} in this work is to experimentally characterize and understand SiMRA's potential for generating PUF responses in commercial off-the-shelf (COTS) DRAM chips. To this end, we first experimentally characterize PUF responses generated using SiMRA in \chipCount{} COTS DDR4 DRAM chips across two major parameters: 1) the number of simultaneously activated rows, and 2) DRAM chip density \& revision. We analyze how temperature affects the similarity of SiMRA-generated responses. Second, we evaluate the quality of the generated PUF responses using standard PUF metrics (i.e., inter- and intra-Jaccard indices~\jaccardDRAMCiteBlock) and evaluation latency for varying numbers of simultaneously activated rows.

Based on our extensive real DRAM chip experiments, we make eight new empirical observations. We summarize our analysis with two key findings. First, we show that SiMRA-PUF provides average intra-Jaccard indices of \actTwoIntraJac, \actFourIntraJac, \actEightIntraJac, \actSixteenIntraJac, and \actThirtyTwoIntraJac, and average inter-Jaccard indices of \actTwoInterJac, \actFourInterJac, \actEightInterJac, \actSixteenInterJac, and \actThirtyTwoInterJac~for 2-, 4-, 8-, 16-, and 32-row activations, respectively. Second, the 2-row activation-based SiMRA-PUF leads to \fracLatencyDiff{} lower evaluation latency than the state-of-the-art COTS DRAM-based PUF~\cite{9923819}.
These results demonstrate that SiMRA-PUF can be used as a fast, low-overhead PUF on real COTS DRAM chips.

We make the following contributions:
\begin{itemize}
    \item We demonstrate the first experimental characterization of signatures generated via simultaneous multiple-row activation (SiMRA) in COTS DRAM chips and the suitability of these signatures as physical unclonable function (PUF) responses.
    \item We characterize \chipCount{} COTS DDR4 DRAM chips under various parameters and conditions in terms of PUF quality: the number of simultaneously activated rows, DRAM chip density \& die revision, and temperature.
    \item We demonstrate that 2-, 4-, 8-, 16-, and 32-row activation-based SiMRA-PUF provides high-quality, reliable, and fast PUF responses. 2-row activation-based SiMRA-PUF leads to \fracLatencyDiff{} lower evaluation latency than the state-of-the-art DRAM-based PUF~\cite{9923819} with the lowest evaluation time.
    \item We open-source our infrastructure at \url{https://github.com/CMU-SAFARI/SiMRA-PUF} to aid future research.
\end{itemize}

%% file: sections/02_background.tex
\section{Background}
\label{sec:background}

\subsection{DRAM Organization and Operation}
\label{sec:dram_org}

\begin{figure}[h]
    \centering
    \includegraphics[width=0.73\linewidth]{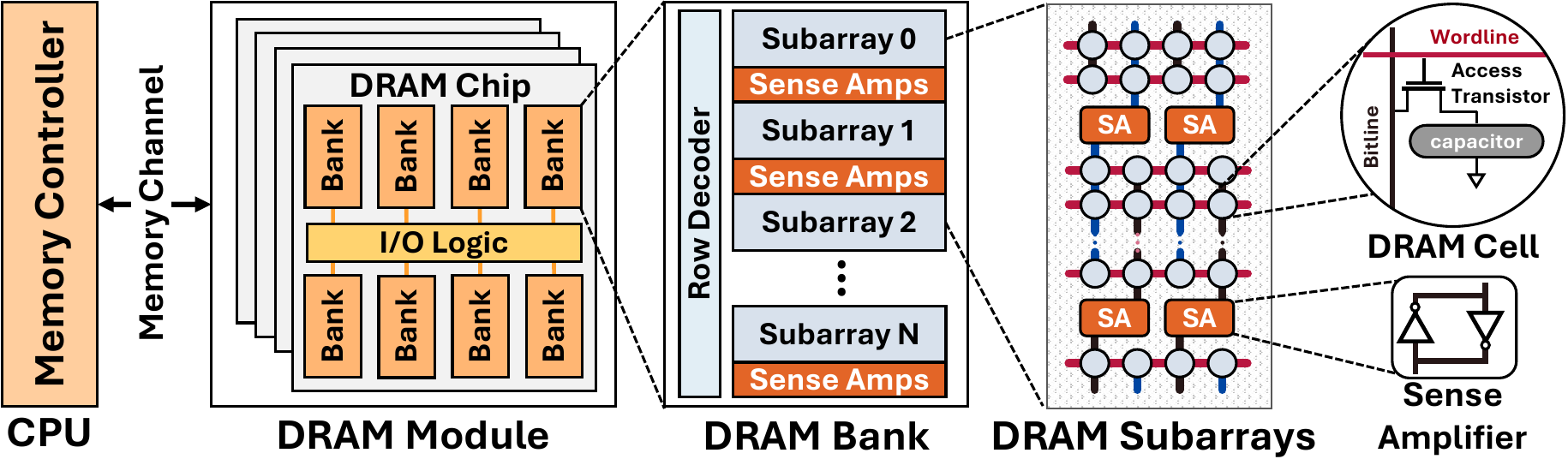}
    \caption{Hierarchical organization of modern DRAM.}
    \label{fig:dram_org}
\end{figure}

Fig.~\ref{fig:dram_org} shows the hierarchical organization of DRAM-based main memory. The memory controller connects to a DRAM module via a memory channel. A module contains one or more ranks, each consisting of multiple DRAM chips operating in lockstep. Each chip contains multiple banks, partitioned into subarrays~\cite{chang2016low, salp, yuksel2025column, seshadri2013rowclone, hajinazar2021simdram, deoliveira2024mimdram}. Within a subarray, DRAM cells are organized in a two-dimensional array of rows (wordlines) and columns (bitlines). A DRAM cell stores one bit of data as electrical charge in a capacitor, accessed via an access transistor driven by a wordline.

The memory controller (MC) accesses DRAM by issuing commands (e.g., \act{}, \pre{}, \rd{}, \wri{})
while respecting a set of timing parameters~\cite{keeth2001dram, lee2013tiered, yuksel2025column, salp,lee2015adaptive, kim2018solar}. To access a row, MC issues an \texttt{ACT} command. The row decoder asserts the corresponding wordline, connecting cells to bitlines. This enables \textit{charge sharing}: the cell shares its charge with the bitline, causing a voltage deviation on the bitline, precharged to $V_{DD}/2$. A sense amplifier
senses this deviation and amplifies the bitline voltage to $V_{DD}$ or $0$~\cite{keeth2001dram, lee2013tiered, yuksel2025column}. To ensure correct operation, MC must wait for charge restoration ($t_{RAS}$) before issuing a \texttt{PRE}, and wait for bitlines to be precharged ($t_{RP}$) before the next \texttt{ACT}~\cite{jedec2017ddr4}.

\subsection{Simultaneous Multiple-Row Activation (SiMRA)}
\label{sec:mra_bg}

Current DRAM standards do \emph{not} officially support activating multiple rows at once. However, the design of COTS DRAM chips does \emph{not} prevent activating multiple (i.e., 2, 4, 8, 16, and 32) DRAM rows at once by issuing an \texttt{ACT$\rightarrow$PRE$\rightarrow$ACT} command sequence 
(called \textit{APA})
with violated $t_{RAS}$ and $t_{RP}$ timing
constraints
\cite{\apaReferences}.

Prior work uses SiMRA in COTS DRAM chips to perform 1) bitwise operations (i.e., MAJ, AND, OR, NOT, NAND, and NOR)~\cite{yuksel2024functionally, gao2019computedram, 9923819, yuksel2024pulsarsimultaneousmanyrowactivation, yuksel2024simultaneous, olgun2022drambender}, 2) data copy and initialization (i.e., RowClone~\cite{seshadri2013rowclone} where one row's content is copied to another row and Multi-RowCopy where one row's content can be copied to up to 31 other different rows simultaneously)~\cite{gao2019computedram,yuksel2024pulsarsimultaneousmanyrowactivation,yuksel2024simultaneous,olgun2022pidram}, 3) true random number generation~\cite{olgun2021quac, yuksel2025in-dram, olgun2022pidram}, and 4) concurrent activation/refresh of two rows in two subarrays~\cite{yaglikci2022hira}.

\subsection{Physical Unclonable Functions (PUFs)}
\label{sec:puf_bg}
PUFs~\cite{gassend2002silicon, Suh2007} map a unique input (i.e., challenge) to a unique signature (i.e., response) by exploiting the physical characteristics of a physical object. The resulting signature reflects the device’s inherent, random physical manufacturing variations. 

Prior work proposes DRAM-based PUFs by violating timing parameters~\cite{kim2018dram, talukder2019prelatpuf, Hashemian:2015, najafi2021,najafi2025epuf,miskelly2020fastdrampuf}, using cell retention failures~\cite{Keller2014, Sutar2016, Xiong2016, schaller2019, liu2014, zheng2021implementation,kumari2018rapid,achievingerrorfree}, storing and sensing cells with fractional voltage levels~\cite{orosa2021codic, 9923819}, exploiting read disturbance bitflips~\cite{schaller:2017, FISCHER2025, anagnostopoulos2018intrinsic, li2023fphammer}, and using startup values~\cite{Tehranipoor:2015,tehranipoor2016dram}. The state-of-the-art DRAM-based PUF generates responses by sensing and amplifying fractional voltage levels stored in cells.

%% file: sections/03_methodology.tex
\section{Experimental Methodology}
\label{sec:methodology}

\subsection{DRAM Characterization Infrastructure}
\label{sec:infra}

\begin{figure}[h]
    \centering
    \includegraphics[width=\linewidth]{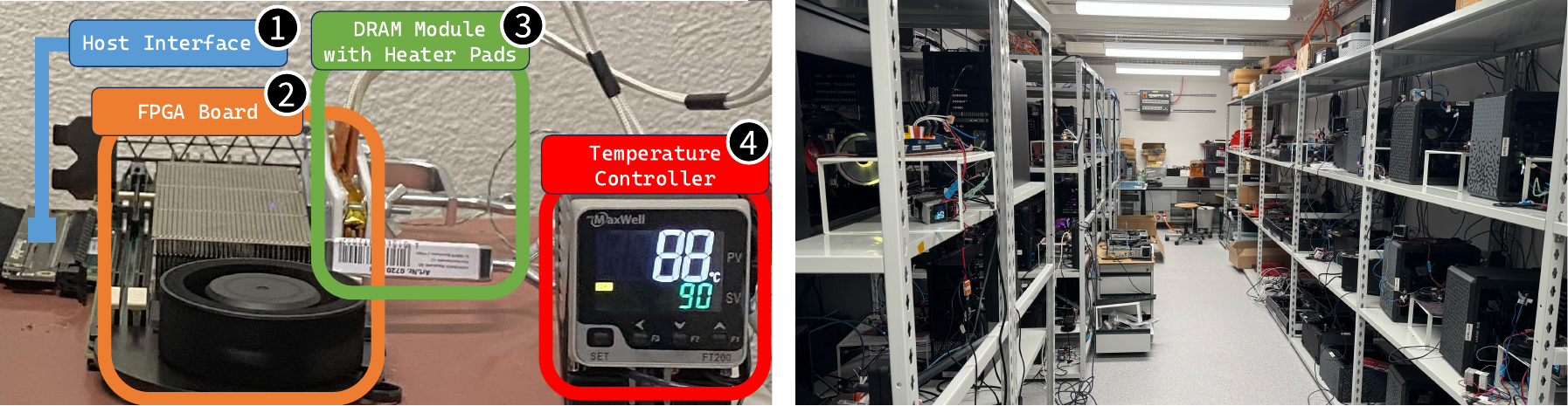}
    \caption{Our DRAM Bender\cite{olgun2022drambender,drambendergithub} based experimental setup (left) and our COTS DRAM chip testing laboratory (right).}
    \label{fig:infra-setup}
\end{figure}

We perform our experiments using DRAM Bender~\cite{olgun2022drambender, drambendergithub} (based on prior SoftMC~\cite{softmcgithub, hassan2017softmc}), an FPGA-based DDR4 DRAM testing infrastructure. \figref{fig:infra-setup} (left) shows our infrastructure that consists of four main components: 1) a host machine that generates DRAM commands, 2) an FPGA board programmed with the DRAM Bender, (Xilinx Alveo U200~\cite{alveo}), 3) a DRAM module with heater pads, and 4) a temperature controller~\cite{maxwellFT200}. This enables us to have i) fine-grained control over DRAM commands and timings, and ii) stable temperature control of the {tested} DRAM chips. \figref{fig:infra-setup} (right) shows our laboratory, which consists of many DDR4 and HBM2 testing platforms.

\subsection{COTS DDR4 DRAM Chips Tested}
\label{sec:dram_tested}
Table~\ref{tab:dram_chips_tested} shows the \chipCount{} DDR4 DRAM chips (\moduleCount~modules) that we focus our analysis on. To investigate whether our characterization study applies to different DRAM technologies, designs, and manufacturing processes, we test a total of 144 DDR4 DRAM chips (14 modules) from all three major manufacturers (i.e., SK hynix, Samsung, Micron). Consistent with prior work \cite{yuksel2025in-dram, yuksel2024functionally, yuksel2024simultaneous, yaglikci2022hira, 9923819, olgun2021quac}, we observe successful SiMRA across \emph{all} tested SK hynix chips, but not in any of other manufacturers' tested chips due to reasons described in~\cite{yuksel2024simultaneous, yaglikci2022hira, yuksel2024functionally}. Thus, we focus our analysis on the SK hynix chips.
\input{tables/00-modules}

\noindent\textbf{Determining DRAM Subarray Boundaries.} To identify rows within the same subarray, we follow the reverse-engineering methodology used in prior works~\cite{seshadri2013rowclone, gao2019computedram, yuksel2025in-dram, yuksel2025column, mutlu2024memory, mutlu2025memorycentric, yuksel2024functionally, olgun2021quac,yuksel2024pulsarsimultaneousmanyrowactivation,yuksel2024simultaneous,yaglikci2022hira,olgun2022pidram,yaglikci2024svard}. We leverage the observation that COTS DRAM chips can copy one row's data (i.e., source row) to another row (i.e., destination row) if and only if the source and destination rows are in the same subarray. We repeatedly perform RowClone for every row pair in a bank. When the destination row successfully updates with the source row's data, we conclude that the two rows are in the same subarray.

\noindent\textbf{Finding Simultaneously Activated Rows.} Prior works~\cite{olgun2021quac, yuksel2025in-dram, yuksel2024functionally, yuksel2024simultaneous, yuksel2024pulsarsimultaneousmanyrowactivation, yuksel2025pudhammer} demonstrate that issuing an \texttt{ACT$\rightarrow$PRE$\rightarrow$ACT} (APA) sequence with significantly reduced $t_{RAS}$ and $t_{RP}$ activates multiple (i.e., 2, 4, 8, 16, or 32) rows simultaneously. To identify these rows, we follow the methodology in~\cite{yuksel2025in-dram}: We issue an APA sequence followed by a \texttt{WR} command with a known data pattern. This \texttt{WR} causes the sense amplifiers to overdrive the bitlines, overwriting the data in all currently open rows. We then read all rows from the bank while adhering to nominal timing parameters. The set of rows containing the written pattern constitutes a \emph{Simultaneously Activated Row (SAR)} group.

\noindent\textbf{Characterization Methodology.}
Our characterization experiment consists of three steps. First, we initialize the rows in the tested SAR group with a predefined data pattern, where each bit of the data pattern corresponds to a row; all cells in that row are initialized to the bit's value. Second, we issue an \textit{APA} command sequence by violating $t_{RAS}$ and $t_{RP}$ timing constraints to activate multiple rows simultaneously. Third, we read the sense amplifier outputs (i.e., the signature). We repeat this experiment 100 times for each SAR group to assess reliability.

\noindent\textbf{Number of SAR Groups Tested.} 
For each tested DRAM module and each activation count (i.e., 2-, 4-, 8-, 16-, and 32-row activation), we uniformly sample 10 SAR groups from each of 30 subarrays in 5 randomly selected banks.

\noindent\textbf{Data Pattern.}
We test \emph{balanced} data patterns (i.e., patterns with an equal number of all-0 and all-1 rows). For each activation count, if the number of balanced patterns is smaller than 100, we test all. Otherwise, we test 100 balanced patterns by including periodic patterns (e.g., \texttt{101010...}~and~\texttt{001100...}) and randomly sampling from the remaining patterns to limit the larger parameter space (e.g., 32-row activation has more than 601M distinct balanced patterns). In total, we test 278 data patterns across all activation counts.

\noindent\textbf{PUF Quality Evaluation Metrics.} We evaluate the quality of SiMRA-generated signatures using the intra- and inter-Jaccard indices, similar to prior work~\jaccardDRAMCiteBlock{}. The Jaccard index~\cite{jaccard1901etude} measures the similarity of two sets ($A$ and $B$), and is defined as $J(A,B) = \frac{|A \cap B|}{|A \cup B|}$. A Jaccard index of 1 indicates that A and B are \textit{identical}, while an index of 0 indicates that A and B are \textit{disjoint}. Thus, an ideal PUF has Jaccard index values of 1 within the same device for the same challenge (i.e., intra-Jaccard) and 0 across different devices for the same challenge (i.e., inter-Jaccard). The intra-Jaccard index represents the stability of a device's responses to the same challenge (higher is better), and inter-Jaccard represents the uniqueness of different devices' responses to the same challenge (lower is better).

\noindent\textbf{Temperature.} To characterize how temperature affects SiMRA-PUF response similarity, we evaluate SiMRA-PUF at 50$^{\circ}$C, 55$^{\circ}$C, 60$^{\circ}$C, 70$^{\circ}$C, and 85$^{\circ}$C. All experiments are conducted at 50$^{\circ}$C, unless stated otherwise.

%% file: tables/00-modules.tex
\begin{table}[h!]
\centering
\vspace{1mm}
\caption{Summary of DDR4 DRAM chips tested.}
\label{tab:dram_chips_tested}
\tiny
\setlength{\tabcolsep}{3pt}
\renewcommand{\arraystretch}{0.5}
\resizebox{0.8\columnwidth}{!}{%
\begin{tabular}{@{}ccccccc@{}}
\toprule
\multirow{2}{*}{{\bf Chip Mfr.}} & {{\bf Module}} & \textbf{\#Modules} & {{\bf Die}} & {{\bf Chip}} & {{\bf Chip}} & {{\bf Date}} \\ 
& {{\bf Mfr.}} & \textbf{(\#Chips)} & {{\bf Rev.}} & {{\bf Density}} & {{\bf Org.}} & {{\bf year-week}} \\ 
\hline \hline \noalign{\vskip 1mm}
\multirow{5}{*}{SK hynix} 
                          & TimeTec     & 2 (16) & A  & 4Gb & x8 & N/A \\ 
                          & TeamGroup   & 2 (16) & M  & 4Gb & x8 & N/A \\
                          & SK hynix    & 3 (32) & A  & 8Gb & x8 & 18-43 \\ 
                          & SK hynix    & 2 (32) & J  & 8Gb & x8 & N/A \\
                          & SK hynix    & 1 (16) & M  & 8Gb & x8 & N/A \\
\bottomrule
\end{tabular}%
}
\end{table}

%% file: sections/04_simra-puf.tex
\section{SiMRA-PUF}\vspace{-0.3mm}

\label{sec:simra-puf}
\noindent\textbf{Key Idea.} The \emph{key idea} of SiMRA-PUF is to generate \emph{device-specific} signatures by exploiting the charge-sharing process that arises from simultaneously activating multiple DRAM rows holding opposing data patterns.

\noindent\textbf{Generating Signatures With SiMRA.} An APA command sequence~(\secref{sec:mra_bg}) simultaneously activates multiple rows initialized with a \emph{balanced} data pattern~(\secref{sec:methodology}), enabling charge sharing between the activated cells and the bitlines. Because cells holding opposing values, the resulting bitline perturbation does not strongly deviate from the reference voltage. However, due to process- and design-induced variations, each bitline settles at a voltage near, but slightly different from, the reference voltage. Then, sense amplifiers resolve bitlines to logic-0 or logic-1. For some bitlines, the perturbation falls below the sense amplifier's reliable sensing margin, and the sense amplifier resolves the bitline randomly as logic-0 or logic-1 across SiMRA trials. For other bitlines, the perturbation exceeds the reliable sensing margin, and the sense amplifier resolves the bitline to a single, stable value (logic-0 or logic-1) across multiple SiMRA trials.

\noindent\textbf{Generating a SiMRA Signature.} To generate a signature, we 1)~initialize a SAR group with a \emph{balanced} data pattern (\secref{sec:methodology}), 2)~issue an APA command sequence (\secref{sec:mra_bg}), and 3)~read the sense amplifier outputs (i.e., the signature). 

\noindent\textbf{Challenge-Response Definition.}
We define each PUF challenge as a (Bank$_{ID}$, Subarray$_{ID}$) pair. 
For each module and each activation count, we select the data pattern that yields the highest number of \emph{good} SAR groups. 
We define a SAR group as \emph{good} if its 64K-bit response contains at least 512 zeros and at least 512 ones (i.e., the response is not biased toward the all-zeros or all-ones pattern). 
For each SAR group, we perform signature generation 100 times and compute the per-bit Shannon entropy~\cite{shannon1948mathematical} across these signatures, averaged over all bits. 
We select the SAR group with the lowest average per-bit entropy (i.e., the SAR group whose bits flip least across repeated signatures) as the representative for that subarray.\footnote{\setstretch{0.8}We conservatively avoid using multiple SAR groups from the same subarray because we observe correlated signatures within a subarray (not shown). With a more detailed analysis, multiple SAR groups from a single subarray can likely be used with SiMRA-PUF, which we provide in an extended version~\cite{simra-puf-extended}.}
The PUF response is the signature read from the selected SAR group.

\subsection{Real DRAM Chip Characterization}
We experimentally characterize SiMRA-PUF on COTS DRAM chips along two parameters: 1)~the number of simultaneously activated rows, and 2)~DRAM chip density \& die revision. We analyze how temperature affects the similarity of SiMRA-generated responses. We also compare SiMRA-PUF against the state-of-the-art DRAM-based PUF, Frac-based PUF~\cite{9923819}, in terms of response quality and evaluation latency.

\begin{figure}[h]
    \centering
    \includegraphics[width=0.75\linewidth]{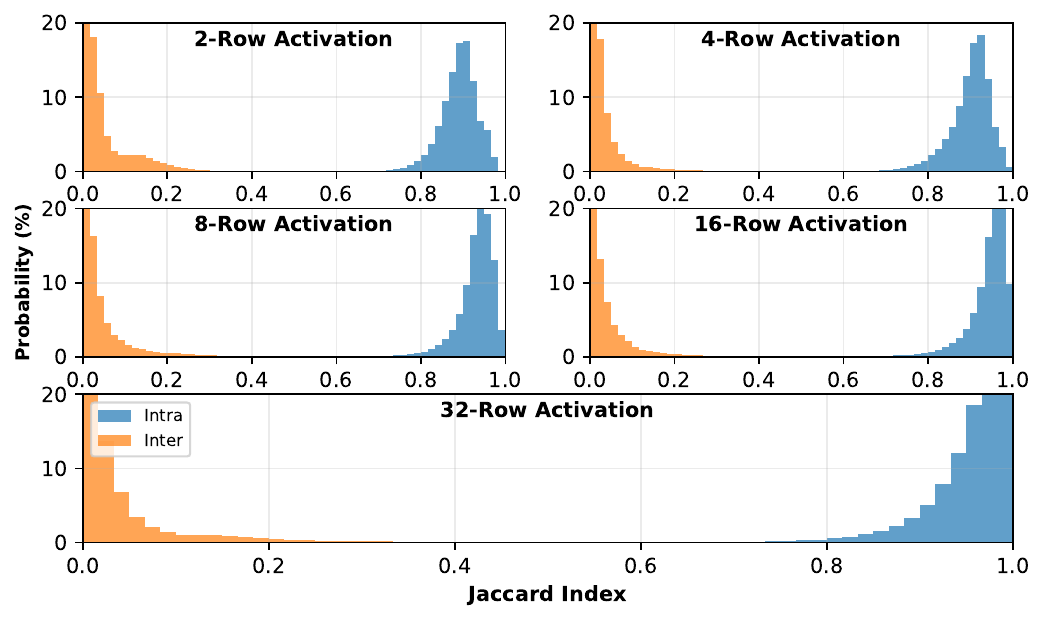}
    \caption{Inter- (orange) and intra-Jaccard (blue) indices obtained for 2/4/8/16/32-row activation-based SiMRA-PUF.}
    \label{fig:simra-puf}
\end{figure}

\noindent\textbf{Effect of the Number of Simultaneously Activated Rows.} \figref{fig:simra-puf} shows the inter- (orange) and intra-Jaccard (blue) indices for 2-, 4-, 8-, 16-, and 32-row activations.

\obsv{SiMRA-PUF generates unique responses across all tested activation counts.}

SiMRA-PUF provides average inter-Jaccard indices of {\actTwoInterJac}, {\actFourInterJac}, {\actEightInterJac}, {\actSixteenInterJac}, and {\actThirtyTwoInterJac} for 2-, 4-, 8-, 16-, and 32-row activations, respectively.

\obsv{SiMRA-PUF responses are highly stable across all tested activation counts.}

SiMRA-PUF provides average intra-Jaccard indices of {\actTwoIntraJac}, {\actFourIntraJac}, {\actEightIntraJac}, {\actSixteenIntraJac}, and {\actThirtyTwoIntraJac} for 2-, 4-, 8-, 16-, and 32-row activations, respectively.

We conclude that SiMRA-PUF produces unique and repeatable responses across all tested activation counts.

\begin{figure}[h]
    \centering
    \includegraphics[width=0.75\linewidth]{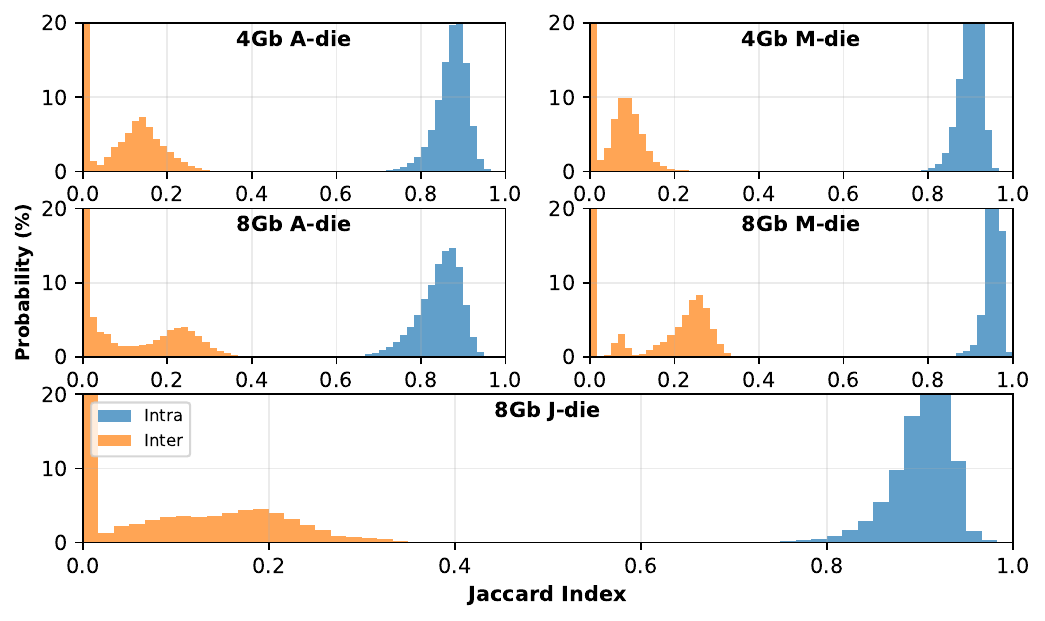}
    \caption{Intra- and inter-Jaccard indices for 2-row activation-based SiMRA-PUF across chip density \& die revisions.}
    \label{fig:simra-puf-act2-per-die}

\end{figure}

\noindent\textbf{Effect of DRAM Chip Density \& Die Revision.}
\figref{fig:simra-puf-act2-per-die} shows the intra- and inter-Jaccard indices of 2-row activation-based SiMRA-PUF across five chip density \& die revision pairs.\footnote{\setstretch{0.8}We observe a similar trend across all SiMRA-PUF designs. We provide a more detailed analysis in the extended version of this paper~\cite{simra-puf-extended}.}

\obsv{DRAM architecture significantly affects the uniqueness of responses.}

We observe that 8Gb M-die and 8Gb J-die DRAM modules exhibit higher inter-Jaccard indices than the other tested modules, indicating that they produce less unique responses.

\obsv{Response stability is consistent across all tested DRAM architectures.}

Intra-Jaccard remains high across all tested DRAM architectures, indicating that all modules produce repeatable responses.

We conclude that DRAM architecture significantly affects response uniqueness but \emph{not} response stability.

\begin{figure}[h]
    \centering
    \includegraphics[width=0.55\linewidth]{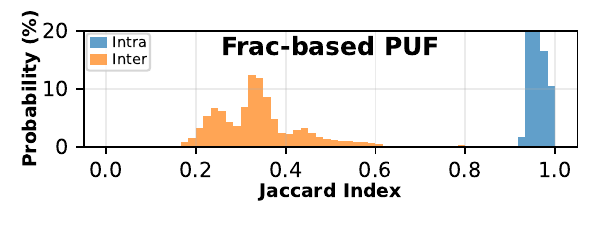}
    \caption{Inter- and intra-Jaccard indices for Frac-based PUF.}
    \label{fig:frac}
\end{figure}

\noindent\textbf{SiMRA-PUF vs. State-of-the-Art DRAM-based PUF.} The state-of-the-art DRAM-based PUF, Frac-based PUF~\cite{9923819} initializes a target row (i.e., the challenge) with all ones and then repeatedly applies the \texttt{Frac} operation (i.e., back-to-back \texttt{ACT$\rightarrow$PRE} command pairs with reduced timing) to drive the DRAM cells toward $V_{DD}/2$; a subsequent activation resolves each cell to logic-0 or logic-1 due to process variation, yielding a device-unique response. \figref{fig:frac} shows the intra- and inter-Jaccard indices of \emph{Frac-based PUF}~\cite{9923819}. For each SAR group selected for 2-row activation-based SiMRA-PUF, we use its first activated row as a Frac-based PUF challenge.\footnote{\setstretch{0.8}We restrict our comparison to intra-Jaccard as Frac-based PUF generates dense responses while SiMRA-PUF generates sparse; a direct comparison of inter-Jaccard indices between different response types would be misleading.}
\vspace{1mm}

\obsv{SiMRA-PUF provides intra-Jaccard indices comparable to those of Frac-based PUF, with the gap narrowing as the number of simultaneously activated rows increases.}

SiMRA-PUF provides average intra-Jaccard indices of \actTwoIntraJac, \actFourIntraJac, \actEightIntraJac, \actSixteenIntraJac, and \actThirtyTwoIntraJac{} for 2-, 4-, 8-, 16-, and 32-row activations, respectively, compared to \fracIntraJac{} for Frac-based PUF. The gap between SiMRA-PUF and Frac-based PUF narrows from \fracGapTwoRow{} at 2-row activation to \fracGapThirtyTwoRow{} at 32-row activation.

\begin{figure}[h]
    \centering
    \includegraphics[width=0.8\linewidth]{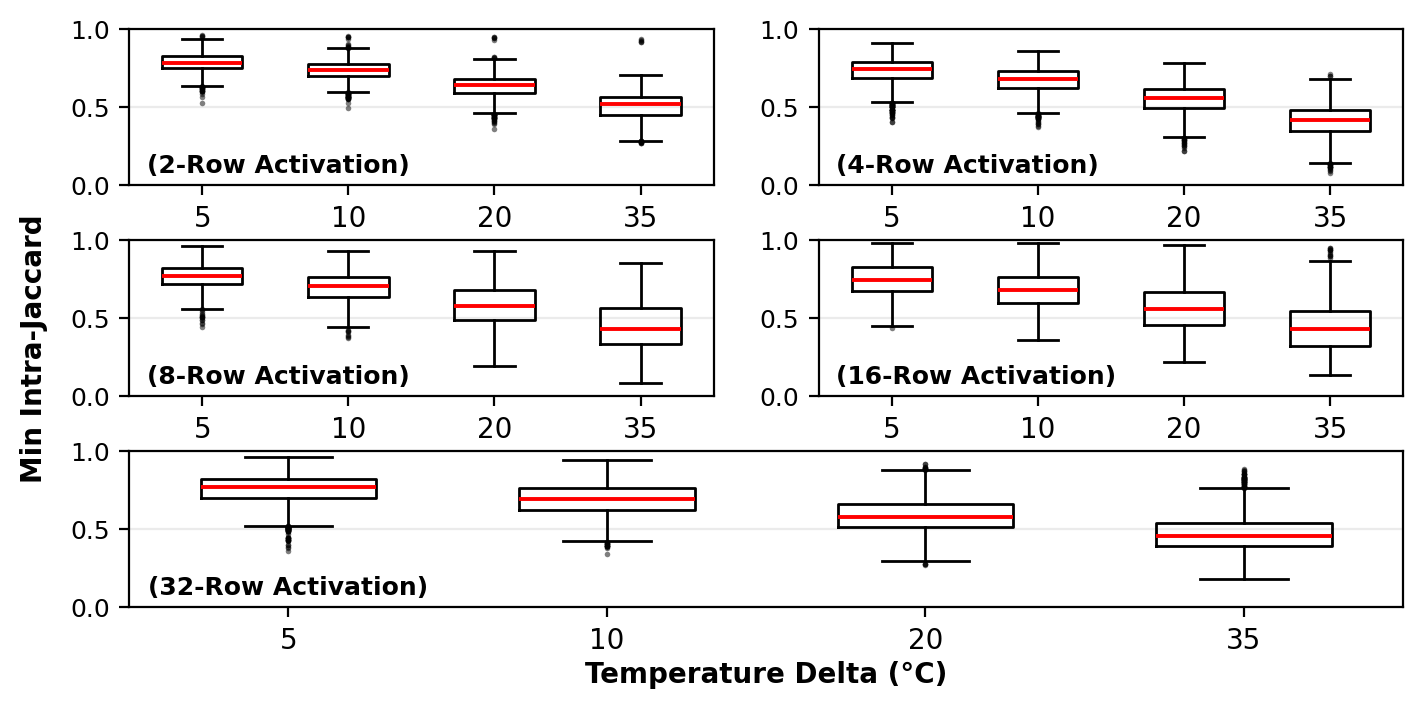}
\caption{Minimum intra-Jaccard indices of SiMRA-PUF at 55-85$^{\circ}$C across 2/4/8/16/32-row activations, vs. the 50$^{\circ}$C baseline.}
    \label{fig:simra-puf-temperature}

\end{figure}

\noindent\textbf{Effect of Temperature.} To evaluate how temperature affects SiMRA-PUF response repeatability, we generate 100 responses for each challenge on the same module at four evaluation temperatures (55$^{\circ}$C, 60$^{\circ}$C, 70$^{\circ}$C, and 85$^{\circ}$C) and compare them against an enrollment baseline collected at 50$^{\circ}$C. \figref{fig:simra-puf-temperature} shows the minimum intra-Jaccard index observed for each challenge across 2-, 4-, 8-, 16-, and 32-row activation-based SiMRA-PUF designs under these four evaluation temperatures.

\obsv{The minimum intra-Jaccard index of every SiMRA-PUF design strictly decreases as the temperature delta between evaluation and enrollment increases.}

We observe that, across all SiMRA-PUF designs, the minimum intra-Jaccard index strictly decreases as the temperature delta increases. Thus, evaluating a SiMRA-PUF at a temperature different from its enrollment temperature (in this case 50$^{\circ}$C) reduces repeatability. To mitigate this effect, we recommend enrolling multiple golden responses at varying temperature intervals, similarly to prior DRAM-based PUF works~\cite{kim2018dram, Sutar2016}.

\obsv{2-row activation-based SiMRA-PUF leads to the best temperature stability among all tested SiMRA-PUF designs; the number of simultaneously activated rows alone does \emph{not} determine SiMRA-PUF temperature stability.}

We observe that 2-row activation leads to the highest minimum intra-Jaccard index on average at every tested temperature. We also observe that the intra-Jaccard drop across the tested temperature range is smallest for 2- and 32-row activations. This behavior shows that the number of simultaneously activated row count alone does \emph{not} determine temperature stability.

We conclude that 1) temperature significantly affects SiMRA-PUF response repeatability and 2) the number of simultaneously activated rows does \emph{not} strongly correlate with SiMRA-PUF temperature stability.

\subsection{PUF Evaluation Latency}

\label{sec:puf-eval}
We evaluate the latency of SiMRA-PUF by measuring the time required for three operations: 1) initializing the SAR group by performing N RowClone operations (one per row), 2) issuing the APA command sequence on the SAR group,\footnote{\setstretch{0.8}For the APA command sequence timings, we use the worst-case successful timings across all tested DRAM modules.} and 3) reading the sense amplifier outputs.\footnote{\setstretch{0.8}For both SiMRA-PUF and Frac-based PUF, we use DDR4-3200W speed bin~\cite{jedec2017ddr4} timings for row read latency.} We evaluate Frac-based PUF~\cite{9923819} as described in the original paper.\footnote{\setstretch{0.8}We provide a more detailed analysis on Frac-based PUF with different numbers of Frac operations and timing parameters on the extended version~\cite{simra-puf-extended}.} Table~\ref{table:latency_PUFs} shows the PUF evaluation latency of Frac-based PUF and SiMRA-PUF with 2-, 4-, 8-, 16-, and 32-row activations.

\obsv{2-row activation-based SiMRA-PUF outperforms Frac-based PUF~\cite{9923819}} in terms of PUF evaluation latency.

We observe that
2-row activation-based SiMRA-PUF has \fracLatencyDiff{} lower evaluation latency than Frac-based PUF. We also observe that SiMRA-PUF's evaluation latency grows significantly with the number of activated rows. For example, 32-row activation-based SiMRA-PUF has an evaluation latency \fracLatencyDiffThirtyTwoRow{} that of Frac-based PUF.

We conclude that 2-row activation-based SiMRA-PUF provides lower latency than the state-of-the-art DRAM-based PUF, while higher activation counts incur significantly higher latency.

\input{tables/01-latency}
\vspace{-0.3mm}

%% file: tables/01-latency.tex
\begin{table}[h]
    \caption{Evaluation time comparison of Frac-based PUF design and SiMRA-PUF designs with 2/4/8/16/32-row activations.}
    \centering
    \footnotesize{}
    \resizebox{\columnwidth}{!}{%
      \setlength{\aboverulesep}{0pt}
  \setlength{\belowrulesep}{0pt}
    \vspace{-1mm}
    \begin{tabular}{ c  c  c }
      \toprule 
        {\vspace{-1mm}\bf Frac-based PUF} & {\bf SiMRA-PUF (2-row)} & {\bf SiMRA-PUF (4-row)} \\
        895 ns & 843.5 ns & 1016.5 ns \vspace{-0.5mm}\\
        \midrule
        {\vspace{-1mm}\bf SiMRA-PUF (8-row)} & {\bf SiMRA-PUF (16-row)} & {\bf SiMRA-PUF (32-row)} \\
        1362.5 ns & 2054.5 ns & 3438.5 ns \vspace{-0.5mm}\\
        \bottomrule
    \end{tabular}%
    }
    \label{table:latency_PUFs}
\end{table}

%% file: sections/05_related-work.tex
\section{Related Work}

\label{sec:related_works}
\vspace{-0.3mm}
To our knowledge, this is the first work to experimentally demonstrate and characterize the use of SiMRA as a PUF substrate in \chipCount~COTS DDR4 DRAM chips. 

\noindent\textbf{DRAM-based PUFs.} 
Prior works exploit DRAM start-up values~\cite{Tehranipoor:2015,tehranipoor2016dram}, cell retention failures~\cite{Keller2014, Sutar2016, Xiong2016, schaller2019, liu2014, zheng2021implementation, kumari2018rapid,achievingerrorfree}, reduced DDRx timing parameters~\cite{kim2018dram, talukder2019prelatpuf, Hashemian:2015, najafi2021,najafi2025epuf,miskelly2020fastdrampuf}, read disturbance bitflips~\cite{schaller:2017, FISCHER2025, anagnostopoulos2018intrinsic, li2023fphammer}, or by storing \& sensing DRAM cells with fractional voltage levels~\cite{orosa2021codic, 9923819} to realize PUFs. \secref{sec:simra-puf} already compares SiMRA-PUF against the state-of-the-art, DRAM-based PUF, Frac-based PUF~\cite{9923819}.

\noindent\textbf{PUFs Based on Other Memory Technologies.}
Several prior works realize PUFs by exploiting unique characteristics of different memory technologies (e.g., SRAM~\cite{Bacha2015, Bhargava2012, holcomb2009power, holcomb2007initial, xiao2014, zheng2013, yue2024srampuf, cortez2012sram, holcomb2014bitline, vijayakumar2017,pandey2016, s2ram}, Flash~\cite{wang2012flash, jiaflash2015, larimian2020, nand-puf}, and emerging memory technologies~\cite{vatajelu2015stt,koeberl2013memristor, rose2013foundations, iyengar2014dwm, das2015,afghah2018reram}).

\noindent\textbf{Multiple-Row Activation in COTS DRAM.} 
Prior works~\cite{\apaReferences} demonstrate that real off-the-shelf DRAM chips can activate multiple rows simultaneously to perform 1) bulk bitwise operations~\cite{yuksel2024functionally, gao2019computedram, 9923819, yuksel2024pulsarsimultaneousmanyrowactivation, yuksel2024simultaneous, olgun2022drambender}, 2) bulk data copy operations~\cite{olgun2022pidram,mutlu2024memory,gao2019computedram,yuksel2024simultaneous}, 3) true random number generation~\cite{olgun2021quac, yuksel2025in-dram, olgun2022pidram}, and 4) concurrent activation/refresh of two rows in two subarrays~\cite{yaglikci2022hira}.

\noindent\textbf{System Integration of COTS DRAM-based Techniques.}
DR-STRaNGe~\cite{bostanci2022dr} provides an end-to-end system design for DRAM-based TRNGs. PiDRAM~\cite{olgun2022pidram} provides a flexible framework for system integration and the evaluation of COTS DRAM-based processing-using-memory techniques. These techniques can be extended to support and evaluate SiMRA-PUF.

\noindent\textbf{Processing-using-DRAM (PuD) in Modified DRAM.}
Prior works~\modifiedPumCiteBlock{} modify the DRAM to enable bulk operations. Similar approaches can improve PuD-based PUFs.

\noindent\textbf{Processing-using-Memory in Other Memory Technologies.} Prior works enable operations by exploiting the operational properties of different memory technologies (e.g., SRAM~\sramPumCiteBlock, Flash~\flashPumCiteBlock, and emerging memory technologies~\emergingPumCiteBlock). These technologies can also benefit from SiMRA-based PUFs.\vspace{-0.3mm}

%% file: sections/06_discussion.tex
\section{Conclusion}
\vspace{-0.3mm}
In this paper, we experimentally demonstrate and characterize, for the first time, the signature-generating capabilities of SiMRA and its usability for generating PUF responses. Through an extensive study using \chipCount~DDR4 DRAM chips (\moduleCount~DRAM modules), we show that i) SiMRA can be used to generate signatures suitable for PUF responses, ii) the number of simultaneously activated rows improves intra-Jaccard and worsens evaluation latency, and iii) temperature significantly affects the similarity of SiMRA-generated PUF responses.\vspace{-0.3mm}

%% file: sections/ack.tex
\section*{Acknowledgments}
\vspace{-0.3mm}We thank the anonymous reviewers of DSN Disrupt 2026. We thank the SAFARI Research Group members for providing a stimulating intellectual and scientific environment. We acknowledge the generous gifts from our industrial partners, including Google, Huawei, Intel, and Microsoft. This work was in part supported by the Google Security and Privacy Research Award and the Microsoft Swiss Joint Research Center.

%% file: sections/07_appendix.tex
\clearpage
\appendix
\nobalance
\onecolumn
\begin{landscape}
\section{\changedText{Appendix: Detailed Per-Module Results.}}
\label{sec:appendix}
Table~\ref{tab:extended-module-info} shows SiMRA-PUF results for each tested DDR4 module (Table~\ref{tab:dram_chips_tested}). We provide two intra-Jaccard (stability) and three inter-Jaccard (uniqueness) metrics per module. For stability, we report the average intra-Jaccard index 1)~across 2-, 4-, 8-, 16-, and 32-row activation-based SiMRA-PUF and 2)~at evaluation temperatures of 55$^{\circ}$C, 60$^{\circ}$C, 70$^{\circ}$C, and 85$^{\circ}$C (averaged across all activation counts) relative to an enrollment baseline collected at 50$^{\circ}$C. For uniqueness, we report three inter-Jaccard indices: \emph{intra-bank} (responses to different challenges within the same bank), \emph{inter-bank} (responses to challenges from different banks of the same module), and \emph{inter-module} (responses to challenges from different modules).
\input{tables/02-modules_extended}
\end{landscape}
\clearpage
\twocolumn

%% file: tables/02-modules_extended.tex
\begin{table}[h]
\centering
\caption{Detailed characteristics and per-module SiMRA-PUF response quality (intra- and inter-Jaccard indices) of all tested DDR4 modules.}
\label{tab:extended-module-info}
\begin{threeparttable}
\resizebox{\ifdim\width>\columnwidth\columnwidth\else\width\fi}{!}{%
\begin{tabular}{ccccccllc|c|c|c|c|c}
\toprule
\multirow{2}{*}{\textbf{\begin{tabular}[c]{@{}c@{}}DRAM\\ Mfr.\end{tabular}}} &
  \multirow{2}{*}{\textbf{\begin{tabular}[c]{@{}c@{}}Module\\ Mfr.\end{tabular}}} &
  \multirow{2}{*}{\textbf{\begin{tabular}[c]{@{}c@{}}Die Density\\ (Gb)\end{tabular}}} &
  \multirow{2}{*}{\textbf{\begin{tabular}[c]{@{}c@{}}Die\\ Rev.\end{tabular}}} &
  \multirow{2}{*}{\textbf{DQ}} &
  \multirow{2}{*}{\textbf{\begin{tabular}[c]{@{}c@{}}Frequency\\ (MT/s)\end{tabular}}} &
  \multicolumn{1}{c}{\multirow{2}{*}{\textbf{\begin{tabular}[c]{@{}c@{}}DRAM\\ Part\end{tabular}}}} &
  \multicolumn{1}{c}{\multirow{2}{*}{\textbf{\begin{tabular}[c]{@{}c@{}}DIMM\\ Part\end{tabular}}}} &
  \multirow{2}{*}{\textbf{\begin{tabular}[c]{@{}c@{}}DIMM Date\\ Code\end{tabular}}} &
  \multicolumn{1}{|c}{\multirow{2}{*}{\textbf{\begin{tabular}[c]{@{}c@{}}Avg. Intra-Jaccard Index\\ (2/4/8/16/32 Row-Activation)\end{tabular}}}} &
  \multicolumn{1}{|c}{\multirow{2}{*}{\textbf{\begin{tabular}[c]{@{}c@{}}Avg. Intra-Jaccard Index\\ ($\Delta$T = 5/10/20/35\,$^{\circ}$C)\textsuperscript{$\dagger$}\end{tabular}}}} &
  \multicolumn{3}{|c}{\textbf{\begin{tabular}[c]{@{}c@{}}Avg. Inter-Jaccard Index\\ (2/4/8/16/32 Row-Activation)\end{tabular}}} \\
\cmidrule(lr){12-14}
 &  &  &  &  &  &  &  &  &  &  & \textbf{Intra-Bank} & \textbf{Inter-Bank} & \textbf{Inter-Module} \\
\midrule
\rowcolor[HTML]{DAE8FC}
SK hynix & TimeTec & 4 & A & x8 & 2133 & H5AN4G8NAFR-TFC
%~\citeMod{H5AN4G8NAFR}
& Unknown & Unknown & 87.40\% / 89.34\% / 94.01\% / 95.35\% / 96.36\% & 73.82\% / 65.39\% / 51.96\% / 38.29\% & 10.07\% / 7.99\% / 18.74\% / 13.08\% / 13.00\% & 9.79\% / 7.78\% / 18.12\% / 12.32\% / 12.72\% & 3.87\% / 1.99\% / 2.39\% / 1.99\% / 2.29\% \\
SK hynix & TimeTec & 4 & A & x8 & 2133 & H5AN4G8NAFR-TFC
%~\citeMod{H5AN4G8NAFR}
& 75TT21NUS1R8-4 & Unknown & 88.00\% / 88.75\% / 93.17 \% /  94.78\% / 95.82\% & 72.71\% / 64.74\% / 51.65\% / 38.44\% & 12.47\% / 5.55\% / 13.59\% / 10.85\% / 9.67\% & 12.10\% / 5.72\% / 13.24\% / 9.86\% / 9.61\% & 4.44\% / 2.32\% / 3.02\% / 2.64\% / 2.46\% \\
\rowcolor[HTML]{DAE8FC}
SK hynix & SK hynix & 4 & A & x8 & 2133 & Unknown & HMA41GU6AFR8N-TF \citeMod{hma41gu6afr8n} & 17-49 & 87.02\% / 89.20\% / 92.97\% / 94.13\% / 94.93\% & 71.85\% / 65.98\% / 54.50\% / 42.49\% & 9.61\% / 5.83\% / 13.40\% / 11.32\% / 14.89\% & 8.81\% / 5.66\% / 12.90\% / 11.04\% / 14.07\% & 3.77\% / 1.85\% / 2.66\% / 2.06\% / 2.73\% \\
SK hynix & SK hynix & 4 & A & x8 & 2133 & Unknown & HMA41GU6AFR8N-TF~\citeMod{hma41gu6afr8n} & 17-49 & 86.68\% / 89.91\% / 93.81\% / 94.40\% / 95.52\% & 74.17\% / 68.06\% / 55.25\% / 42.87\% & 10.36\% / 8.47\% / 17.39\% / 13.83\% / 16.14\% & 9.96\% / 8.26\% / 16.67\% / 13.37\% / 14.75\% & 3.53\% / 1.98\% / 2.65\% / 1.96\% / 2.78\% \\
\rowcolor[HTML]{DAE8FC}
SK hynix & TeamGroup & 4 & M & x8 & 2666 & H5AN4G8NMFR-TFC
%~\citeMod{H5AN4G8NMFR}
& TLRD44G2666HC18F-SBK~\citeMod{TLRD44G2666HC18F} & Unknown & 89.62\% / 90.31\% / 92.51\% / 93.57\% / 92.68\% & 75.08\% / 71.16\% / 61.31\% / 47.72\% & 4.62\% / 3.12\% / 3.45\% / 3.08\% / 5.83\% & 4.78\% / 3.22\% / 3.31\% / 3.11\% / 5.54\% & 2.69\% / 1.84\% / 1.95\% / 1.72\% / 1.95\% \\
SK hynix & TeamGroup & 4 & M & x8 & 2666 & H5AN4G8NMFR-TFC
%~\citeMod{H5AN4G8NMFR}
& TLRD44G2666HC18F-SBK~\citeMod{TLRD44G2666HC18F} & Unknown & 90.27\% / 90.37\% / 91.69\% / 92.74\% / 92.85\% & 71.17\% / 66.82\% / 58.91\% / 47.71\% & 5.25\% / 3.70\% / 3.26\% / 3.52\% / 4.47\% & 5.47\% / 3.65\% / 3.00\% / 3.45\% / 4.09\% & 2.73\% / 1.81\% / 1.87\% / 1.76\% / 1.62\% \\
\rowcolor[HTML]{DAE8FC}
SK hynix & SK hynix & 8 & A & x8 & 2400 & H5AN8G8NAFR-UHC
%~\citeMod{H5AN8G8NAFR}
& HMA81GU7AFR8N-UH~\citeMod{hma8182gu} & 18-43 & 84.31\% / 87.58\% / 88.01\% / 89.42\% / 92.29\% & 75.19\% / 69.42\% / 58.89\% / 46.21\% & 8.62\% / 12.59\% / 11.38\% / 10.84\% / 17.52\% & 8.53\% / 11.84\% / 10.12\% / 10.67\% / 15.58\% & 2.43\% / 2.13\% / 3.26\% / 2.31\% / 1.77\% \\
SK hynix & SK hynix & 8 & J & x8 & 2666 & Unknown & HMA82GU6JJR8N-VK~\citeMod{HMA82GU6JJR8N} & Unknown & 90.40\% / 86.80\% / 92.70\% / 94.70\% / 95.36\% & 74.94\% / 69.07\% / 57.80\% / 44.71\% & 9.08\% / 1.49\% / 6.77\% / 11.95\% / 19.70\% & 8.90\% / 1.48\% / 6.84\% / 11.60\% / 17.43\% & 3.26\% / 0.91\% / 2.70\% / 2.24\% / 1.91\% \\
\rowcolor[HTML]{DAE8FC}
SK hynix & TimeTec & 8 & J & x8 & 2133 & H5AN8G8NJJR-VKC & Unknown & Unknown & 89.98\% / 90.84\% / 94.80\% / 95.98\% / 97.27\% & 83.57\% / 78.62\% / 70.64\% / 60.37\% & 9.43\% / 9.62\% / 18.17\% / 13.77\% / 25.71\% & 9.03\% / 8.95\% / 17.29\% / 13.92\% / 24.87\% & 3.18\% / 2.35\% / 3.44\% / 2.54\% / 2.26\% \\
SK hynix & SK hynix & 8 & M & x8 & 2133 & Unknown & HMA82GS6MFR8N-TF~\citeMod{eightGigMDie} & Unknown & 95.34\% / 94.80\% / 96.75\% / 95.59\% / 95.54\% & 82.50\% / 74.92\% / 62.05\% / 47.49\% & 12.60\% / 9.80\% / 10.15\% / 11.00\% / 14.00\% & 11.77\% / 9.27\% / 10.01\% / 10.12\% / 12.66\% & 3.97\% / 1.68\% / 1.78\% / 1.73\% / 1.43\% \\
\bottomrule
\end{tabular}%
}
\begin{tablenotes}[flushleft]
\footnotesize
\item[$\dagger$] The $\Delta$T intra-Jaccard index is averaged across all
activation counts (i.e., 2/4/8/16/32-row).
\end{tablenotes}
\end{threeparttable}
\end{table}